\documentclass[11pt]{article}
\usepackage{amsmath}
\begin{document}
\title {\Large\bf Homotopy Structure of 5d Vacua}
\vskip 1.0cm

\author{Eun Kyung Park\\
{\small The Liberal Arts Course, Kyungsung University, Pusan
608-736, Korea}\\ \\
 and \\ \\
 Pyung Seong Kwon\footnote{E-mail:bskwon@star.ks.ac.kr}\\
 {\small Department of Physics, Kyungsung University, Pusan
608-736, Korea}}

\vspace{1.4cm}

\date{}

\maketitle \thispagestyle{empty} \baselineskip 6.5mm

 It is shown that flat zero-energy solutions (vacua) of the 5d Kaluza-Klein theory admit a non-trivial
homotopy structure generated by certain Kaluza-Klein excitations.
These vacua consist of an infinite set of homotopically different
spacetimes denoted by $\mathcal{M}^{(n)}_5$, among which
$\mathcal{M}^{(0)}_5$ and $\mathcal{M}^{(1)}_5$ are especially
identified as $M_{4} \times S^{1}$ and $M_5$, the vacuum states of
the 5d Kaluza-Klein theory and the 5d general relativity,
respectively (where $M_k$ represents the $k$-dimensional Minkowski
space).

\vskip 2cm
\medskip
\begin{center}
\end{center}

\newpage
\baselineskip 6.0mm

\setcounter{page}{1}
  The 5d Kaluza-Klein theory is distinguished from the ordinary 5d general relativity by
the fact that the background vacuum is assumed to be the product
$M_{4} \times S^{1}$ instead of $M_5$, where $M_k$ represents the
$k$-dimensional Minkowski space. Having zero energy, the manifolds
$M_4 \times S^1$ and $M_5$ are both qualified for the vacuum state
of the 5d theory of gravity. Classically, there is no way to
determine which is more appropriate than the other. The only basis
for choosing $M_4 \times S^1$ as the vacuum state of the 5d
Kaluza-Klein theory is that it includes the compactified dimension
which is crucial in order to admit an internal gauge group $U(1)$
in the reduced theory. In fact, the compactification is an
essential ingredient of any higher dimensional theory (including
string theory) based on the Kaluza-Klein theory. The particle
spectrum is then obtained by expanding around the vacuum $M_4
\times S^1$ ; one finds a finite number of massless modes, and an
infinite tower of massive (excitation) modes. In traditional
theories the low-energy physics would be mostly governed by the
dynamics of the massless modes alone, because the energy scale of
the massive modes is about of order of the Planck scale. However,
this is not to be the case anymore once we adopt the scenario that
extra dimensions be very large. Recently, it has been suggested
that the old hierarchy problem can be solved in the framework of
higher-dimensional theories by taking extra dimensions to be very
large\cite{1}. In this scenario the energy scale of the massive
modes (which is of order the inverse of the radius of $S^1$) could
sufficiently lower down to the level of low energy physics, and
one can imagine that the massive modes perhaps play an important
role even in the low energy limit. In this paper, we examine the
effect of the massive excitations on the geometry or topology of
the background spacetime $M_4 \times S^1$. Then we end up with a
remarkable result that 5d vacua admit a non-trivial homotopy
structure ; 5d vacua consist of an infinite set of homotopically
different spacetimes denoted by $\mathcal{M}^{(n)}_5$, and where
$\mathcal{M}^{(0)}_5$ and $\mathcal{M}^{(1)}_5$ are especially
identified as $M_4 \times S^1$ and $M_5$, respectively.

We start the discussion with a metric
\begin{eqnarray}
ds^2 &=& -dt^2 + dr^2 + r^2 (d\theta^2 + \sin^2\theta d\phi^2) +
\Phi^2 (x^{\alpha}, x^5)
[dx^5 + A_{\mu} (x^{\alpha}, x^5)dx^{\mu}]^2 \nonumber \\
&=& {\mathop{g}\limits^4}_{\mu\nu} dx^{\mu} dx^{\nu} + \Phi^2
[dx^5 + A_{\mu} dx^{\mu}]^2 \nonumber \\
&=& {\mathop{\eta}\limits^4}_{ab} \omega^a \omega^b + (\omega^5)^2
\,\,, \,\,\,\,\,\,\,\,\,\,(\mu, \nu, \alpha = 0, 1, 2, 3 \,\,;\,\,
a, b = 0, 1, 2, 3)\,\,,
\end{eqnarray}
where ${\mathop{\eta}\limits^4}_{ab}$ is the 4d flat Minkowski
metric, and
\begin{equation}
\hspace{-3.0cm}\omega^a = e^a\,_\mu dx^{\mu} \,\,,
\,\,\,\,\,\,\,\,\,\,\, \omega^5 = \Phi (dx^5 + A_{\mu} dx^{\mu})
\,\,\,\,\,\,\,\,\,\,
\end{equation}
are basis 1-forms of the orthonormal frame. From Eq.(1) we see
that the fields $\Phi$ and $A_\mu$ have $x^5$-dependence, meaning
that they include massive modes. In the reduced 4d sector, $\Phi$
and $A_{\mu}$ are identified as the Brans-Dicke scalar and the
$U(1)$ gauge potential, respectively. The vierbein $e^a\,_\mu$, on
the other hand, are functions of $x^{\alpha}$ alone\footnote{The
$x^5$-independency of $e^a\,_\mu$ greatly simplifies the
calculation. See ref.\cite{2}}, so they commute with $\partial_5$.
The non-vanishing components of Riemann tensor are then calculated
in the orthonormal frame to give\footnote{Note that the
constraints $\mathcal{D}_{[a}{\hat{f}_{bc]}} - F_{[a}
\hat{f}_{bc]} =0$ and $\mathcal{D}_{[a} F_{b]} + (\partial_5
\hat{f}_{ab}) / 2\Phi = 0$ are required by the symmetry
${\mathop{R}\limits^5}_{ABCD}= {\mathop{R}\limits^5}_{CDAB}$, and
they have been used to obtain ${\mathop{R}\limits^5}_{abc5}$ and
${\mathop{R}\limits^5}_{a5b5}$. Also, note that the 4d sector of
the spacetime in Eq.(1) is flat, so the 4d Riemann tensor
${\mathop{R}\limits^4}_{abcd}$ vanishes and does not appear in
Eq.(3). The Riemann tensors in Eqs.(3) to (5) are in fact a
generalization of those in ref.\cite{3}.}
\begin{equation}
\hspace{-1.3cm}{\mathop{R}\limits^5}_{abcd} = -\frac{1}{2} (\hat
{f}_{ab} \hat{f}_{cd} + \hat{f}_{_{_{_{_{\scriptstyle a}}}}\,[c}
\hat{f}_{_{_{\scriptstyle b}}\,d]}) \,\,,
\end{equation}
\begin{equation}
\hspace{-2.6cm} {\mathop{R}\limits^5}_{abc5} =
\mathcal{D}_{[a}{\hat{f}_{b]c}} - F_{c} \hat{f}_{ab} \,\,,
\end{equation}
\begin{equation}
\hspace{-0.6cm} {\mathop{R}\limits^5}_{a5b5}= - \mathcal{D}_{(a}
F_{b)} -\frac{1}{4} \hat{f}_{ac}\hat{f}^{c}\,\,_{b} - F_{a}F_{b}
\,\,,
\end{equation}
where $\hat{f}_{ab}$ is defined by $\hat{f}_{ab} \equiv \Phi
f_{ab}$ and
\begin{equation}
f_{ab} = e_{a}\,^{\mu} e_{b}\,^{\nu} (D_{\mu} A_{\nu} - D_{\nu}
A_{\mu}) \equiv e_{a}\,^{\mu} e_{b}\,^{\nu} f_{\mu\nu}
\,\,,\,\,\,\,\,\,\,\,(D_{\mu} \equiv \partial_{\mu} - A_{\mu}
\partial_5 )\,\,,
\end{equation}
\begin{equation}
F_a = e_{a}\,^{\mu} (\partial_{\mu} \Phi -
\partial_5 \hat{A}_{\mu}) / \Phi \equiv e_{a}\,^{\mu} F_{\mu}
\,\,,\,\,\,\,\,\,\,\,(\hat{A}_{\mu} \equiv \Phi A_{\mu}) \,\,,
\end{equation}
\begin{equation}
\mathcal{D}_a = e_{a}\,^{\mu} (\nabla_{\mu} - A_{\mu} \partial_5)
\equiv e_{a}\,^{\mu} \mathcal{D}_{\mu}\,\,\,\,,
\end{equation}
and in Eq.(8), $\nabla_{\mu}$ represents the ordinary covariant
derivative associated with the metric
${\mathop{g}\limits^4}_{\mu\nu}$. Also, $f_{\mu\nu}$ in Eq.(6) is
a generalization of the Maxwell field strength ; it takes the same
form as the conventional Maxwell field strength except that the
ordinary derivative $\partial_{\mu}$ is replaced by the covariant
derivative\footnote{The derivative $D_{\mu}$ (more generally
$\mathcal{D}_{\mu}$), acting on a tensor, leaves its components
invariant under the gauge transformation: $x^5 \rightarrow
{x^5}^{\prime} = x^5 + f(x^{\alpha})$, $A_{\mu} \rightarrow
A_{\mu}^{\prime} = A_{\mu} -
\partial_{\mu}f$. See ref.\cite{4} for this. Thus we see that the Eqs.(3), (4)
and (5) are all expressed in gauge-covariant form.} $D_{\mu}$.
Without $x^5$-dependency the derivative $D_{\mu}$ reduces to
$\partial_{\mu}$, and consequently $f_{\mu\nu}$ becomes the
conventional Maxwell field strength.

Now we look for flat (vacuum) solutions which satisfy the
equations
${\mathop{R}\limits^5}_{abcd}={\mathop{R}\limits^5}_{abc5}=
{\mathop{R}\limits^5}_{a5b5}=0$. A set of the simplest solutions
to these equations may be obtained by setting
\begin{equation}
f_{\mu\nu} = D_{\mu} A_{\nu} - D_{\nu} A_{\mu} = 0 \,\,,
\end{equation}
\begin{equation}
F_{\mu} = \frac{1}{\Phi}(\partial_{\mu} \Phi -
\partial_5 \hat{A}_{\mu}) = 0 \,\,,
\end{equation}
and in particular Eq.(9) is immediately solved by an ansatz
\begin{equation}
A_t = A_{\theta} = A_{\phi}=0 \,\,, \,\,\,\,\,\,\,\,\,\,\, A_r =
A_r(r, x^5) \,\,, \,\,\,\,\,\,\,\,\,\,\, \Phi=\Phi (r, x^5) \,\,.
\end{equation}
Equation (11) is actually the most general ansatz preserving
spherical symmetry, and $A_{\mu}$ in Eq.(11) may be regarded as a
pure gauge in the sense that it gives $f_{\mu\nu}=0$. With this
ansatz, the metric in Eq.(1) can be recast into the form
\begin{equation}
ds^2 = -dt^2 + d\hat{r}^2 + r^2 (d\theta^2 + \sin^2\theta d\phi^2)
+ \frac{\Phi^2}{1+ \hat{A}_r\,^2}
(dx^5)^2\,\,,\,\,\,\,\,\,\,(\hat{A}_r \equiv \Phi A_r)\,\,,
\end{equation}
where $d\hat{r}$ is defined by
\begin{equation}
d\hat{r} = (1+ \hat{A}_r\,^2)^{1/2} dr + \frac{\Phi \hat{A}_r}{(1+
\hat{A}_r\,^2)^{1/2}}dx^5 \,\,.
\end{equation}
Equation (13) indicates that the variable $\hat{r}$ is a function
of $r$ and $x^5$ ; i.e., $\hat{r} = f(r, x^5)$ with
\begin{equation}
\frac{\partial f}{\partial r} = (1+ \hat{A}_r\,^2)^{1/2} \,\,,
\,\,\,\,\,\,\,\frac{\partial f}{\partial x^5} = \frac{\Phi
\hat{A}_r}{(1+ \hat{A}_r\,^2)^{1/2}} \,\,.
\end{equation}
With the aid of Eq.(10) one then finds from the equations in (14)
that the condition $\partial^2 f/ \partial r \partial x^5 =
\partial^2 f/
\partial x^5 \partial r$ implies that $\partial_r \hat{A}_r = 0$ ;
i.e., $\hat{A}_r$ must be a function of $x^5$ alone:
\begin{equation}
\hat{A}_r \equiv Y (x^5) \,\,.
\end{equation}
Thus the equations in (14) are now integrated to give
\begin{equation}
f = \hat{r} =  r (1+ Y^2)^{1/2} + g(x^5) \,\,,
\end{equation}
\begin{equation}
\Phi = r Y^{\prime} + \frac{(1+ Y^2)^{1/2}}{Y} g^{\prime}\,\,,
\end{equation}
where $g(x^5)$, which has been introduced as an integral constant,
is an arbitrary function of $x^5$ alone, and the 'prime' in
Eq.(17) denotes the $x^5$-derivative. Note that $\hat{A}_r$ and
$\Phi$ in Eqs.(15) and (17) indeed describe the flat solution
satisfying ${\mathop{R}\limits^5}_{ABCD}=0$ ; one can readily
check that they satisfy Eq.(10). Now we impose the condition
\begin{equation}
\lim_{Y, Y^{\prime}  \rightarrow 0} \Phi = \rm{constant} \equiv
\Phi_0 \,\,,
\end{equation}
which suggests that the solution we are to find is the one that
reduces to the Kaluza-Klein vacuum $M_4 \times S^1$ as $A_{\mu}
\rightarrow 0$ ; note that $M_4 \times S^1$ with $A_{\mu}=0$ and
$\Phi = \rm{constant}$ is also a solution to Eq.(10). The
condition in Eq.(18) immediately implies that
\begin{equation}
g^{\prime} = \Phi_0 Y \,\,,
\end{equation}
and therefore $\Phi$ in Eq.(17) becomes
\begin{equation}
\Phi = r Y^{\prime} + \Phi_0 (1+ Y^2 )^{1/2} \,\,.
\end{equation}
Using all this, one can show that the metric in Eq.(12) can be
converted into the form
\begin{equation}
ds^2 = -dt^2 + (1-\frac{\Phi^2_0 Y^2}{R^2})d\rho^2 +
\frac{\rho^2}{1+Y^2}(d\theta^2 + \sin^2\theta d\phi^2) + R^2 (dx^5
+ \frac{\Phi_0 Y}{R^2}d\rho)^2\,\,,
\end{equation}
where $\rho$ and $R$ are defined by
\begin{equation}
\rho = \hat{r} - g = r(1+ Y^2)^{1/2}\,\,,
\end{equation}
\begin{equation}
R = [\Phi^2_0 Y^2 + (\rho \frac{Y^{\prime}}{1+Y^2} + \Phi_0 )^2
]^{1/2} \,\,.
\end{equation}

So far, the function $Y(x^5)$ and the constant $\Phi_0$ have been
entirely arbitrary except that they should satisfy the condition
(18). Now let us take
\begin{equation}
Y_n (x^5) = \tan \frac{n x^5}{2R_c}
\,\,,\,\,\,\,\,\,\,\,\,\,\,\,\,\,\,\,\Phi_0 =
\delta_{n0}\,\,,\,\,\,\,\,\,\,\,\,\,\,\,(n=0,1,2,\cdots)\,\,,
\end{equation}
where $R_c$ represents the compactification radius of the
fifth-dimension (note that Eq.(24) respects the condition (18) as
$n \rightarrow 0$). By Eqs.(15), (20) and (24), the gauge field
$A_r$ becomes
\begin{equation}
A_r (r,x^5) = a_n (r) \sin \frac{nx^5}{R_c} \,\,
\end{equation}
with
\begin{equation}
a_n (r) = \frac{R_c}{nr + 2 R_c \delta_{n0}} \,\,,
\end{equation}
which shows that taking $Y_n (x^5)$ as in Eq.(24) implies that we
are considering a situation where the n-th excitation of the gauge
field (together with the scalar field induced by this gauge field)
is present in the background spacetime $M_4 \times S^1$ (see
Eq.(1)). In this case the field $\Phi (r,x^5)$ takes the form
\begin{equation}
\Phi (r, x^5) = \frac{nr}{2R_c} \sec^2 \frac{nx^5}{2R_c} +
\delta_{n0} \,\,,
\end{equation}
and we see that $A_r \rightarrow 0$, $\Phi \rightarrow 1$ for
$n=0$. Namely, the $n=0$ state simply describes the compactified
vacuum $M_4 \times S^1$ without any gauge, or scalar field. For $n
\neq 0$, on the other hand, both fields are present in the form of
$A_r \sim 1/r$ and $\Phi \sim r$, respectively (the presence of
the gauge excitation necessarily demands the presence of the
scalar field with a behavior of $\Phi \sim r$, as can be checked
from Eq.(10)). Then, what happens to the spacetime by the presence
of these excitations $\Phi$ and $A_r$? By Eq.(24), the metric (21)
simplifies to
\begin{equation}
ds^2 = -dt^2 + d\rho^2 + \rho^2 \sin^2 \chi_{_n} (d\theta^2 +
\sin^2\theta d\phi^2) + (\rho d\chi_{_n} + \delta_{n0} dx^5)^2\,\,
\end{equation}
with $\chi_{_n}$ defined by
\begin{equation}
\chi_{_n} (x^5) = \frac{nx^5}{2R_c} + \frac{\pi}{2} \,\,,
\end{equation}
which, after all, implies that the metric (28) is equivalent under
(24) to the metric in Eq.(1) with $\Phi$ and $A_\mu$ given by
Eqs.(25) and (27). Indeed, for $n=0$, it reduces to the flat $M_4
\times S^1$ :
\begin{equation}
ds^2_{(0)} = -dt^2 + d\rho^2 + \rho^2 (d\theta^2 + \sin^2\theta
d\phi^2) + (dx^5)^2\,\,,
\end{equation}
which is just the metric in Eq.(1) with $A_{\mu}=0, \Phi =1$. For
$n \neq 0$, on the other hand, the metric (28) becomes
\begin{equation}
ds^2_{(n)} = -dt^2 + d\rho^2 + \rho^2 [ d\chi^2_{_n} + \sin^2
\chi_{_n} (d\theta^2 + \sin^2\theta d\phi^2)] \,\,,
\end{equation}
the 5d flat Minkowski metric! The topology of spacetime has been
changed by the non-zero excitation modes. The toroidal
compactification $M_4 \times S^1$ in Eq.(1) has been converted
into the non-compact spacetime described by the metric in Eq.(31).
Furthermore, let us change the variable $x^5 \rightarrow y_n$ by
the equation
\begin{equation}
y_n = x^5 + \frac{\pi R_c}{n} \,\,.
\end{equation}
Under Eq.(32), the metric (1) (with $A_r$ and $\Phi$ given by
Eqs.(25) and (27)) takes the same form as before : i.e.,
\begin{equation}
ds^2 = -dt^2 + dr^2 + r^2 (d\theta^2 + \sin^2\theta d\phi^2) +
\Phi^2 (r, y_n) [dy_n + A_r (r, y_n)dr]^2
\end{equation}
with\footnote{The metric (33) seems to have a singularity at $y_n
= 0$ (or $2\pi R_c$) because $\Phi (r,y_n)$ in (34) is singular
there. However, this singularity is obviously artificial ; notice
that the metric (33) is a flat solution satisfying $R_{ABCD}=0$ in
orthonormal frame.}
\begin{equation}
A_r (r,y_n) = - a_n (r) \sin \frac{ny_n}{R_c}
\,\,,\,\,\,\,\,\,\,\, \Phi(r, y_n) = \frac{nr}{2R_c} \csc^2
\frac{ny_n}{2R_c} + \delta_{n0} \,\,,
\end{equation}
but $\chi_{_n}$ in (31) is now written (from (29)) as
\begin{equation}
\chi_{_n} (y) = \frac{ny_n}{2R_c} \,\,, \,\,\,\,\,\,\,\,\,\, ( 0
\leq y_n \leq 2\pi R_c ) \,\,.
\end{equation}
This is remarkable. Equation (35) suggests that the 5d vacua admit
a non-trivial homotopy structure. Note that the angle $\chi_{_n}$
varies from $0$ to $n\pi$ as $y_n$ makes a single turn from $0$ to
$2\pi R_c$ around $S^1$, which in turn means that the set of
variables ($\chi_{_n}, \theta, \phi$) in (31) covers the
hypersurface $S^3$ $n$ times when ($y_n, \theta, \phi$) in (33)
covers $S^2 \times S^1$ just once. The 3d manifold described by
($\chi_n, \theta, \phi$) can be regarded as a 3-loop $\alpha_n
(t_1, t_2, t_3)$, where $t_i$, the coordinates of 3d cube $I_3$,
are defined by the map : $t_1 = y_n /2\pi R_c$, $t_2 = \theta
/\pi$ and $t_3 = \phi /2\pi$, which is an homeomorphism $f : S^2
\times S^1 \longrightarrow I_3$. A collection of these 3-loops
which cover $S^3$ $n$ times constitutes the $n$-th equivalence
class of the 3rd. homotopy group $\bf \pi_3$$(S^3)$. The spatial
subsector of the spacetime described by Eq.(31) is essentially a
pile of such 3-loops. This suggests that the spacetime described
by Eq.(31) (let us call it $\mathcal{M}^{(n)}_5$) belongs to the
$n \neq 0$ homotopy class of $\bf \pi_3$$(S^3)$, though $M_4
\times S^1$ with no excitations especially belongs to the $n=0$
class since it is obtained by simply taking\footnote{Eq.(35)
(i.e., the variable change $x^5 \rightarrow y_n$ in Eq.(32)) is in
fact applicable to the $n=0$ case either. Note that since
$\underset{n \rightarrow 0}\lim \, ny_n = \pi R_c$ from (32), we
see that the metric (33) reduces to the flat $M_4 \times S^1$ as
$n \rightarrow 0$. This is the same result that we obtain from the
metric (1) with $A_r$ and $\Phi$ given by Eqs.(25) and (27).}
$n=0$. Further, for $n=1$, it is obvious from Eq.(35) that
$\mathcal{M}^{(n)}_5$ is precisely identified as the ordinary 5d
Minkowski space $M_5$. But note that, in general,
$\mathcal{M}^{(n)}_5$ is not $M_5$ itself ; $\mathcal{M}^{(n)}_5$
is an $n$-fold cover of $M_5$. Namely, it is a fiber bundle over
$M_5$ with fiber $F$ a discrete set of $n$ points. In short, 5d
vacua consist of an infinite set of homotopically different
spacetimes $\mathcal{M}^{(n)}_5$, among which the cases of $n=0$
and $n=1$ are especially identified as the background vacua of the
5d Kaluza-Klein theory and the 5d general relativity, respectively
; i.e., $M_4 \times S^1 = \mathcal{M}^{(0)}_5$, and $M_5 =
\mathcal{M}^{(1)}_5 $. Such a homotopy structure manifests itself
once there is a defect, or a point particle at $\rho=0$. With a
defect or point particle at $\rho=0$, the spatial subsector of
$M_5$ is not simply connected, and the 3-loops which contain
$\rho=0$ are not shrinkable. So each $\mathcal{M}^{(n)}_5$, a pile
of such 3-loops which enclose the point $\rho=0$ $n$ times,
belongs to a different homotopy class.

Though the above discussion has its own right in 5d Kaluza-Klein
theories, it may be applied to any other higher dimensional
theories with toroidal compactification.  For instance, in the
eleven-dimensional theory compactified on $X \times S^1 / Z_2$ (or
equivalently, in the strong coupling limit of the $E_8 \times E_8$
heterotic string) it is believed that the radius of the orbifold
$S^1$ is larger than the volume of the Calabi-Yau manifold $X$,
and there is a regime where our spacetime appears
five-dimensional\cite{5,6}. To the lowest order of the 11d Newton
constant $\kappa$, the (5d sector of the) ground state metric
takes the form of the 5d Kaluza-Klein vacuum $M_4 \times S^1$. So
in this case the vacua of the theory could admit the homotopy
structure under discussion. If this is the case, it then follows
that the T-duality could break down due to the presence of the
excitations. Note that for $ n \neq 0$ the compactified dimension
$S^1$ disappears due to the excitation modes ; the map $f$ : $S^2
\times S^1 \rightarrow S^3$ takes $S^1$ to the great circle of
$S^3$. Thus the winding number (of the closed string) is not a
topological number here and consequently the term corresponding to
winding modes does not exist in the mass formula, and it leads to
the conjecture that the T-duality might break down in the presence
of excitations.

So far, we have considered only the (excitations of the)
Kaluza-Klein components (i.e., 5d metric components) as the source
of the topology change of the spacetime. But in the 5d bulk
spacetime there also exist other fields besides the Kaluza-Klein
components\cite{6}. For instance, we may consider the case where a
5d $U(1)$ gauge field\footnote{Or it could be the remnant of
dimensionally reduced higher-rank tensor fields.} $\mathcal
{A}_{\mu}$ exists in the bulk spacetime, with field strength
$\mathcal {F}_{\mu\nu}$ (see for instance ref.\cite{7}), and it
takes the place of the Kaluza-Klein vector $A_{\mu}$. Indeed,
disregarding the components $\mathcal {F}_{\mu 5}$, one finds that
the 4d effective action for $\mathcal{A}_{\mu}$ takes the same
form as the action for $A_{\mu}$, which suggests that the homotopy
structure of the 5d vacua can be generated even by (the
excitations of) bulk fields, rather than Kaluza-Klein components.

The result of this paper is quite analogous to the case of the
$\theta$-vacua of the Yang-Mill theory. In both cases the vacua
admit the same homotopy structure with an infinite set of homotopy
classes each of which is characterized by an integer $n$. In the
Yang-Mill theory the integer $n$ is identified with the Pontryagin
index $q \sim Tr \int d^4 x F^{\mu\nu}
{\mathop{F}\limits^\sim}_{\mu\nu} $, and the vacua belonging to
different homotopy classes are connected by a Euclidean
(instanton) solution. In the present paper the integer $n$ is
simply an excitation number of the pure gauge $A_{\mu}$, and the
instanton solution interconnecting two different vacua does not
exist here. But notice that in our case there exist gauge
transformations which mix the massive and massless modes\cite{8}
of $A_\mu$, and consequently the spacetimes belonging to different
classes can be mixed by such gauge transformations. The gauge
transformation which mixes the massive and massless modes can be
generated by allowing gauge parameter $\xi$ to depend on $x^5$.
For instance, the transformation which mixes the $n-$th and
$(n-k)-$th modes (in the Fourier exponential series) of $A_\mu$
takes the form (see ref.\cite{8})
\begin{equation}
\delta_k A_{\mu n} = \delta_{nk} \partial_{\mu}\xi_k + i
(n-2k)\xi_k A_{\mu (n-k)}/R_c \,\,\,\,,
\end{equation}
where $A_{\mu n}$ and $\xi_n$ are the $n-$th components of the
Fourier exponential series
\begin{equation}
A_{\mu}(x^{\alpha}, x^5) = \sum_n A_{\mu n} (x^{\alpha}) e^{inx^5
/R_c} \,\,\,\,,
\end{equation}
\begin{equation}
 \xi (x^{\alpha}, x^5)= \sum_n \xi_n (x^{\alpha}) e^{i
  nx^5 /R_c} \,\,\,\,,
\end{equation}
respectively, and $\delta_k$ represents the transformation induced
by $\xi_k$. Since $\mathcal{M}^{(n)}_5$ is equivalent to $\mathcal
{M}^{(0)}_5$ plus the $n-$th excitation (in the Fourier
sine-series) of $A_{\mu}$, the transformation (36) apparently
mixes $\mathcal{M}^{(n)}_5$ with other spacetimes belonging to
different classes. This is quite remarkable. Note that a gauge
transformation essentially does not change the physics of a
system. But the above argument suggests that $\mathcal{M}^{(n)}_5$
is not invariant under a 'large' gauge transformation of the form
(36), which immediately leads us to suspect that spacetimes
$\mathcal{M}^{(n)}_5$ (most importantly $\mathcal{M}^{(0)}_5 = M_4
\times S^1$ and $\mathcal{M}^{(1)}_5 = M_5$) may not be the
physical states of the theory. Namely, the suggested conjecture is
that $M_4 \times S^1$ may not be the true physical vacuum of the
5d Kaluza-Klein theory ; it serves as a vacuum only when we do not
consider the 'large' gauge transformation. The physical vacuum of
the 5d Kaluza-Klein theory may perhaps be a superposition of an
infinite number of $\mathcal{M}^{(n)}_5$, analogously to the case
of the $\theta-$vacua of the Yang-Mill theory.

 \vskip 2cm

 This research was supported by Kyungsung University
research grants in 2003.

\end{document}